\documentclass[preprint,aps,12pt,showpacs,nofootinbib,tightenlines]{revtex4}
\usepackage{amsmath}
\usepackage{amssymb}
\usepackage{epsfig}
\usepackage{graphicx}
\begin{document}
\def\be{\begin{eqnarray}}
\def\en{\end{eqnarray}}
\def\non{\nonumber\\}

\def\ra{\rangle}
\def\la{\langle}
\def\sl{\!\!\!\slash}
\def\prd{{Phys. Rev. D}~}
\def\prl{{ Phys. Rev. Lett.}~}
\def\plb{{ Phys. Lett. B}~}
\def\npb{{ Nucl. Phys. B}~}
\def\epjc{{ Eur. Phys. J. C}~}
\newcommand{\acp}{{\cal A}_{CP}}

\newcommand{\psl}{ P \hspace{-2.4truemm}/ }
\newcommand{\nsl}{ n \hspace{-2.2truemm}/ }
\newcommand{\vsl}{ v \hspace{-2.2truemm}/ }
\newcommand{\epsl}{\epsilon \hspace{-1.8truemm}/\,  }
\title{Study of $f_0(980)$ and $f_0(1500)$ from $ B_s \to f_0(980)\pi, f_0(1500)\pi$ Decays}
\author{Zhi-Qing Zhang
\footnote{Electronic address: zhangzhiqing@haut.edu.cn} } 
\affiliation{\it \small  Department of Physics, Henan University of Technology,
Zhengzhou, Henan 450052, P.R.China }
\date{\today}
\begin{abstract}
In this paper, we analyze the scalar mesons $f_0(980)$ and
$f_0(1500)$ from the decays $\bar B^0_s \to f_0(980)\pi^0,
f_0(1500)\pi^0$ within Perturbative QCD approach. From the leading order calculations, we find that (a)
in the allowed mixing angle ranges, the branching ratio of $\bar
B^0_s\to f_0(980)\pi^0$ is about $(1.0\sim1.6)\times 10^{-7}$, which
is smaller than that of $\bar B^0_s\to f_0(980)K^0$ (the difference
is a few times even one order); (b) the decay $\bar B^0_s \to
f_0(1500)\pi^0$ is better to distinguish between the lowest lying
state or the first excited state for $f_0(1500)$, because the
branching ratios for two scenarios have about one-order difference
in most of the mixing angle ranges; and (c) the direct CP
asymmetries of $\bar B^0_s \to f_0(1500)\pi^0$ for two scenarios
also exists great difference. In scenario II, the variation range of
the value ${\cal A} ^{dir}_{CP}(\bar B^0_s \to f_0(1500)\pi^0)$
according to the mixing angle is very small,
except for the values corresponding to the mixing angles being
near $90^\circ$ or $270^\circ$, while the variation range of ${\cal A}
^{dir}_{CP}(\bar B^0_s \to f_0(1500)\pi^0)$ in scenario I is very
large. Compared with the future data for the decay $\bar B^0_s \to
f_0(1500)\pi^0$, it is ease to determine the nature of the scalar
meson $f_0(1500)$.
\end{abstract}

\vspace{1cm}

\maketitle


\section{Introduction}\label{intro}
For the  underlying structure of the scalar mesons is still under
controversy, there are two typical schemes for the classification to
them \cite{nato,jaffe}. The nonet mesons below 1 GeV, including
$f_0(600), f_0(980), K^*(800)$ and $a_0(980)$, are usually viewed as
the lowest lying $q\bar q$ states, while the nonet ones near 1.5
GeV, including $f_0(1370), f_0(1500)/f_0(1700), K^*(1430)$ and
$a_0(1450)$, are suggested as the first excited states. Here we
denote this scheme as scenario I, and the following scheme as
scenario II: the nonet mesons near 1.5 GeV are treated as $q\bar q$
ground states, while the nonet mesons below 1 GeV are exotic states
beyond the quark model such as four-quark bound states. In order to
uncover the inner structures, many approaches are used to research
the modes of $B_{u,d}$ decaying into a scalar and a pseudoscalar
(vector) meson, such as the generalized factorization approach
\cite{GMM}, QCD factorization approach (QCDF)
\cite{CYf0K,CCYscalar,CCYvector}, Perturbative QCD (PQCD) approach
\cite{Chenf0K1,Chenf0K2,wwang,ylshen,zqzhang1,zqzhang2}. On the experimental side, along with
the running of the Large Hadron Collider beauty experiments (LHC-b), some of $B^0_s$ decays involved a
scalar in the final states might be observed in the Large Hadron Collider beauty experiments (LHC-b) \cite{lhc1,lhc2}.
In order to make precision studies of rare decays in the B-meson systems, the LHC-b
detector is designed to exploit the large number of b hadrons produced. Furthermore, it can reconstruct a B-decay
vertex with very good resolution, which is essential for studying the rapidly oscillating $B_s$ mesons.
Some of $B^0_s$ decays involved a scalar
in the final states can also serve as an
ideal platform to probe the natures of these scalar mesons. So the
studies of these decay modes for $B^0_s$ are necessary in the next a
few years.

 In this paper, we will study the branching ratios and the direct
CP asymmetries of $\bar B^0_s \to f_0(980)\pi, f_0(1500)\pi$ within Perturbative
QCD approach based on $k_T$ factorization. It is organized as
follows: In Sect.\ref{proper}, we introduce the input parameters including the decay constants
and light-cone distribution amplitudes.  In Sec.\ref{results}, we
then apply PQCD approach to calculate analytically the branching
ratios and CP asymmetries for our considered decays. The final part
contains our numerical results and discussions.


\section{Input Parameters}\label{proper}
In order to make quantitative predictions, we identify
$f_0(980)$ as a mixture of $s\bar s$ and $n\bar n=(u\bar u+d\bar
d)/\sqrt2$, that is
 \be |f_0(980)\ra = |s\bar s\ra\cos\theta+|n\bar
n\ra\sin\theta, \en where the mixing angle $\theta$ is taken in the
ranges of $25^\circ< \theta <40^\circ$ and
$140^\circ<\theta<165^\circ$ \cite{hycheng}. Certainly, $f_0(1500)$
can be treated as a $q\bar q$ state in both scenario I and II. We
consider that the meson $f_0(1500)$ and $f_0(980)$ have the same
component structure but with different mixing angle.

For the the neutral scalar meson $f_0(980), f_0(1500)$
cannot be produced via the vector current, we have $\langle f_0(p)|\bar q_2\gamma_\mu q_1|0\ra=0$.
Taking the mixing into account, the scalar current $\langle f_0(p)|\bar q_2q_1|0\ra=m_S\bar {f_S}$
can be written as:
\be
\langle f_0^n|d\bar
d|0\ra=\langle f_0^n|u\bar u|0\ra=\frac{1}{\sqrt 2}m_{f_0}\tilde
f^n_{f_0},\,\,\,\, \langle f_0^s|s\bar s|0\ra=m_{f_0}\tilde
f^s_{f_0},
\en
where $f_0^{(n,s)}$ represent for the quark flavor states for $n\bar{n}$ and $s\bar{s}$ components of $f_0$ meson,
respectively. For the scalar decay constants $\tilde f_{f_0}^n$ and $\tilde f_{f_0}^s$
are very close\cite{CCYscalar}, we can assume $\tilde
f_{f_0}^n=\tilde f_{f_0}^s$ and denote them as $\bar f_{f_0}$ in the
following.

The twist-2 and twist-3 light-cone distribution amplitudes (LCDAs)
for different components of $f_0$ are defined by:
\be
\langle f_0(p)|\bar q(z)_l q(0)_j|0\rangle
&=&\frac{1}{\sqrt{2N_c}}\int^1_0dxe^{ixp\cdot z}\{p\sl\Phi_{f_0}(x)
+m_{f_0}\Phi^S_{f_0}(x)+m_{f_0}(n\sl_+n\sl_--1)\Phi^{T}_{f_0}(x)\}_{jl},\non
\label{LCDA}
\en
where we assume $f_0^n(p)$ and $f_0^s(p)$ are same and denote
them as $f_0(p)$, $n_+$ and $n_-$ are light-like vectors:
$n_+=(1,0,0_T),n_-=(0,1,0_T)$. The normalization can be related to
the decay constants:
\be \int^1_0 dx\Phi_{f_0}(x)=\int^1_0
dx\Phi^{T}_{f_0}(x)=0,\,\,\,\,\,\,\,\int^1_0
dx\Phi^{S}_{f_0}(x)=\frac{\bar f_{f_0}}{2\sqrt{2N_c}}.
\en

The wave function for $\pi$ meson is
given as \cite{pball}
\be
\Phi_{\pi}(P,x,\zeta)\equiv
\frac{1}{\sqrt{2N_C}}\gamma_5 \left [ \psl \Phi_{\pi}^{A}(x)+m_0^{\pi}
\Phi_{\pi}^{P}(x)+\zeta m_0^{\pi} (\vsl \nsl - v\cdot
n)\Phi_{\pi}^{T}(x)\right ] . \en
where $P$ and $x$ are the momentum and the momentum fraction of
$\pi$ meson, respectively. The parameter $\zeta$ is either $+1$ or $-1$ depending on the
assignment of the momentum fraction $x$.

In general, the $B_s$ meson is treated as heavy-light system and its Lorentz structure
can be written as\cite{grozin,kawa}
\be
\Phi_{B_s}=\frac{1}{\sqrt{2N_c}}(\psl_{B_s}+M_{B_s})\gamma_5\phi_{B_s}(k_1).
\en
For the contribution of $\bar \phi_{B_s}$ is numerically small \cite{caidianlv} and has been neglected.
\section{Theoretical Framework and perturbative calculations} \label{results}
Under the two-quark model for the scalar mesons supposition, we
would like to use PQCD approach to study $\bar B^0_s \to f_0(980)\pi, f_0(1500)\pi$ decays.
In this approach, the decay amplitude is
separated into soft, hard, and harder dynamics characterized by
different energy scales $(t, m_{B_s}, M_W)$. It is conceptually
written as the convolution,
\be
{\cal A}(\bar B^0_s \to f_0\pi)\sim \int\!\!
d^4k_1 d^4k_2 d^4k_3\ \mathrm{Tr} \left [ C(t) \Phi_{B_s}(k_1)
\Phi_{f_0}(k_2) \Phi_{\pi}(k_3) H(k_1,k_2,k_3, t) \right ],
\label{eq:con1}
\en
where $k_i$'s are momenta of anti-quarks included in each mesons, and $\mathrm{Tr}$ denotes the trace over
Dirac and color indices. $C(t)$ is the Wilson coefficient which
results from the radiative corrections at short distance. The function
$H(k_1,k_2,k_3,t)$ describes the four quark operator and the
spectator quark connected by a hard gluon whose $q^2$ is in the order
of $\bar{\Lambda} M_{B_s}$, and includes the $\mathcal{O}(\sqrt{\bar{\Lambda} M_{B_s}})$ hard dynamics.
Therefore, this hard part $H$ can be perturbatively calculated.

Since the $b$ quark is rather heavy, we consider the $\bar B^0_s$ meson at rest
for simplicity. It is convenient to use light-cone coordinate $(p^+,
p^-, {\bf p}_T)$ to describe the meson's momenta,
\be
p^\pm =\frac{1}{\sqrt{2}} (p^0 \pm p^3), \quad {\rm and} \quad {\bf p}_T =
(p^1, p^2).
\en
Using these coordinates the $\bar B^0_s$ meson and the two
final state meson momenta can be written as
\be P_{B_s} =
\frac{M_{B_s}}{\sqrt{2}} (1,1,{\bf 0}_T), \quad P_{2} =
\frac{M_{B_s}}{\sqrt{2}}(1,0,{\bf 0}_T), \quad P_{3} =
\frac{M_{B_s}}{\sqrt{2}} (0,1,{\bf 0}_T), \en respectively. The
meson masses have been neglected. Putting the anti-quark momenta in
$\bar B^0_s$, $f_0$ and $\pi^0$ mesons as $k_1$, $k_2$, and $k_3$, respectively, we
can choose
\be
k_1 = (x_1 P_1^+,0,{\bf k}_{1T}), \quad k_2 = (x_2
P_2^+,0,{\bf k}_{2T}), \quad k_3 = (0, x_3 P_3^-,{\bf k}_{3T}).
\en
For our considered decay channels, the integration over $k_1^-$,
$k_2^-$, and $k_3^+$ in eq.(\ref{eq:con1}) will lead to
\be {\cal
A}(\bar B^0_s \to f_0\pi^0) &\sim &\int\!\! d x_1 d x_2 d x_3 b_1 d b_1 b_2 d
b_2 b_3 d b_3 \non && \cdot \mathrm{Tr} \left [ C(t)
\Phi_{B_s}(x_1,b_1) \Phi_{f_0}(x_2,b_2) \Phi_{\pi}(x_3, b_3) H(x_i,
b_i, t) S_t(x_i)\, e^{-S(t)} \right ], \quad \label{eq:a2} \en where
$b_i$ is the conjugate space coordinate of $k_{iT}$, and $t$ is the
largest energy scale in function $H(x_i,b_i,t)$. The large logarithms ($\ln
m_W/t$) coming from QCD radiative corrections to four-quark
operators are included in the Wilson coefficients $C(t)$. The large
double logarithms ($\ln^2 x_i$) on the longitudinal direction are
summed by the threshold resummation \cite{li02}, and they lead to
$S_t(x_i)$, which smears the end-point singularities on $x_i$. The
last term, $e^{-S(t)}$, is the Sudakov form factor, which suppresses
the soft dynamics effectively \cite{soft}. Thus it makes the
perturbative calculation of the hard part $H$ applicable at
intermediate scale, i.e., $M_{B_s}$ scale.

We will calculate analytically the function $H(x_i,b_i,t)$ for $\bar B^0_s
\to f_0\pi^0$ decays in the leading-order and give the convoluted
amplitudes. For our considered decays, the related weak effective
Hamiltonian $H_{eff}$ can be written as \cite{buras96}
\be
\label{eq:heff} {\cal H}_{eff} = \frac{G_{F}} {\sqrt{2}} \,
\sum_{q=u,c}V_{qb} V_{qs}^*\left[ \left (C_1(\mu) O_1^q(\mu) +
C_2(\mu) O_2^q(\mu) \right)+ \sum_{i=3}^{10} C_{i}(\mu) \,O_i(\mu)
\right] \; ,
\en
with the Fermi constant $G_{F}=1.166 39\times
10^{-5} GeV^{-2}$, and the CKM matrix elements V. We specify below
the operators in ${\cal H}_{eff}$ for $b \to s$ transition:
\be
\begin{array}{llllll}
O_1^{u} & = &  \bar s_\alpha\gamma^\mu L u_\beta\cdot \bar
u_\beta\gamma_\mu L b_\alpha\ , &O_2^{u} & = &\bar
s_\alpha\gamma^\mu L u_\alpha\cdot \bar
u_\beta\gamma_\mu L b_\beta\ , \\
O_3 & = & \bar s_\alpha\gamma^\mu L b_\alpha\cdot \sum_{q'}\bar
 q_\beta'\gamma_\mu L q_\beta'\ ,   &
O_4 & = & \bar s_\alpha\gamma^\mu L b_\beta\cdot \sum_{q'}\bar
q_\beta'\gamma_\mu L q_\alpha'\ , \\
O_5 & = & \bar s_\alpha\gamma^\mu L b_\alpha\cdot \sum_{q'}\bar
q_\beta'\gamma_\mu R q_\beta'\ ,   & O_6 & = & \bar
s_\alpha\gamma^\mu L b_\beta\cdot \sum_{q'}\bar
q_\beta'\gamma_\mu R q_\alpha'\ , \\
O_7 & = & \frac{3}{2}\bar s_\alpha\gamma^\mu L b_\alpha\cdot
\sum_{q'}e_{q'}\bar q_\beta'\gamma_\mu R q_\beta'\ ,   & O_8 & = &
\frac{3}{2}\bar s_\alpha\gamma^\mu L b_\beta\cdot
\sum_{q'}e_{q'}\bar q_\beta'\gamma_\mu R q_\alpha'\ , \\
O_9 & = & \frac{3}{2}\bar s_\alpha\gamma^\mu L b_\alpha\cdot
\sum_{q'}e_{q'}\bar q_\beta'\gamma_\mu L q_\beta'\ ,   & O_{10} & =
& \frac{3}{2}\bar s_\alpha\gamma^\mu L b_\beta\cdot
\sum_{q'}e_{q'}\bar q_\beta'\gamma_\mu L q_\alpha'\ ,
\label{eq:operators} \end{array}
\en
where $\alpha$ and $\beta$ are
the $SU(3)$ color indices; $L$ and $R$ are the left- and
right-handed projection operators with $L=(1 - \gamma_5)$, $R= (1 +
\gamma_5)$. The sum over $q'$ runs over the quark fields that are
active at the scale $\mu=O(m_{B_s})$, i.e.,
$(q'\epsilon\{u,d,s,c,b\})$.

\begin{figure}[t,b]
\vspace{-4cm} \centerline{\epsfxsize=20 cm \epsffile{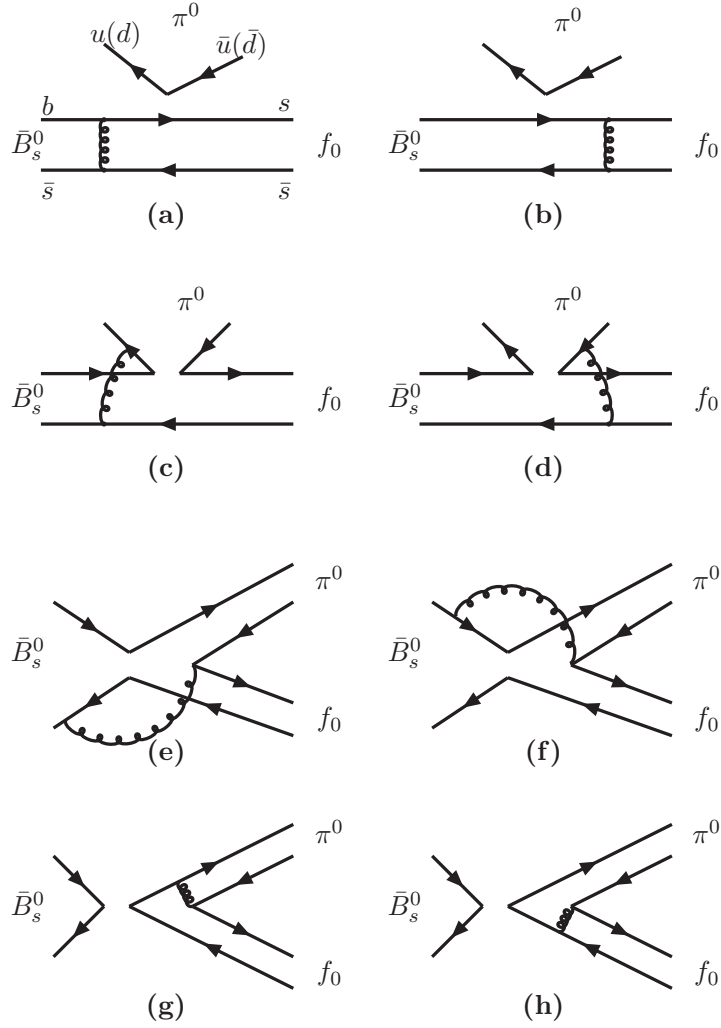}}
\vspace{-10.5cm} \caption{ Diagrams contributing to the $\bar B^0_s\to
f_0\pi^0$ decays.}
 \label{Figure1}
\end{figure}

We will show the whole amplitude for each diagram including wave
functions. There are 8 type diagrams contributing to the $\bar B^0_s \to
f_0\pi^0$ decays are illustrated in Fig.\ref{Figure1}. We first
calculate the usual factorizable diagrams (a) and (b). Operators
$O_{1,2,3,4,9,10}$ are $(V-A)(V-A)$ currents, and the operators
$O_{5,6,7,8}$ have the structure of $(V-A)(V+A)$, the sum of the their
amplitudes are written as $F_{ef_0}$ and $F_{ef_0}^{P1}$,
respectively.
\begin{eqnarray}
F_{ef_0} &=&F^{P1}_{ef_0}= 8 \pi C_F m_{B_s}^4 f_{\pi}\int_0^1 dx_1 dx_2
\int_0^{\infty} b_1db_1\, b_2db_2\, \Phi_{B_s}(x_1,b_1)
\nonumber \\
& & \times \left\{ \left[ (1+x_2)\Phi_{f_0}(x_2)-r_{f_0}(1-2x_2)
\left( \Phi_{f_0}^S(x_2)+\Phi_{f_0}^T(x_2) \right) \right]
\right.\nonumber\\
& & \left. \times E_{ei}(t) h_{e}(x_1,x_2,b_1,b_2) -2r_{f_0}
\Phi_{f_0}^S({x_2})E_{ei}(t') h_{e}(x_2,x_1,b_2,b_1)
\right\},\end{eqnarray}
where $f_{\pi}$ is the decay constant of $\pi$ meson, $r_{f_0}=m_{f_0}/m_{B_s}$.

In some other cases, we need to do Fierz transformation  for the corresponding
operators to get right flavor and color structure for factorization to work. We may get
$(S-P)(S+P)$ operators from $(V-A)(V+A)$ ones. For
these $(S-P)(S+P)$ operators, Fig.~1(a) and 1(b) give

\begin{eqnarray}
 F^{P2}_{ef_0}&=&16 \pi C_F
m_{B_s}^4 f_{\pi}r_{\pi}\int_0^1 dx_1 dx_2 \int_0^{\infty} b_1db_1\,
b_2db_2\, \Phi_{B_s}(x_1,b_1)
\nonumber \\
& &\times \left\{ -\left[ \Phi_{f_0}(x_2)+r_{f_0} \left(
x_2\Phi_{f_0}^T(x_2)-(x_2+2)\Phi_{f_0}^S(x_2) \right) \right]
\right.\nonumber\\
& &\left.\times E_{ei}(t) h_{e}(x_1,x_2,b_1,b_2)+2r_{f_0}
\Phi_{f_0}^S({x_2})E_{ei}(t') h_{e}(x_2,x_1,b_2,b_1) \right\}
\;, \end{eqnarray}
where $r_{\pi}=m^{\pi}_0/m_{B_s}$.

For the non-factorizable diagrams 1(c) and 1(d), all three meson
wave functions are involved. The integration of $b_2$ can be
performed using $\delta$ function $\delta(b_3-b_2)$, leaving only
integration of $b_1$ and $b_3$. Here we have two kinds of
contributions: $M_{ef_0}$, $M_{ef_0}^{P1}$ and $M_{ef_0}^{P2}$
describe the contributions from the $(V-A)(V-A)$, $(V-A)(V+A)$ and $(S-P)(S+P)$ operators, respectively.
\begin{eqnarray}
{\cal M}_{ef_0} &=& 32 \pi C_Fm_{B_s}^4 /\sqrt{2N_C}\int_0^1
dx_1dx_2dx_3 \int_0^{\infty} b_1 db_1\, b_3
db_3\,\Phi_{B_s}(x_1,b_1) \Phi_{\pi}^A(x_3)
\nonumber \\
&&\times
 \bigg\{-[(x_3-1)\Phi_{f_0}(x_2)-r_{f_0}x_2(\Phi_{f_0}^S(x_2)-\Phi_{f_0}^T(x_2))]
E'_{ei}(t) h_n(x_1,1-x_3,x_2,b_1,b_3)
\nonumber \\
&&-\left[(x_2+x_3)\Phi_{f_0}(x_2) +r_{f_0} x_2
(\Phi_{f_0}^S(x_2)+\Phi_{f_0}^T(x_2))\right] E'_{ei}(t')
h_n(x_1,x_3,x_2,b_1,b_3) \bigg\},\;\;\quad
\end{eqnarray}
\begin{eqnarray}
{\cal M}^{P1}_{ef_0} &=& 32 \pi C_Fm_{B_s}^4 /\sqrt{2N_C}r_{\pi}\int_0^1
dx_1dx_2dx_3 \int_0^{\infty} b_1 db_1\, b_3
db_3\,
\nonumber \\
&&\times \Phi_{B_s}(x_1,b_1)
\bigg\{[(x_3-1)\Phi_{f_0}(x_2)(\Phi_{\pi}^P(x_3)+\Phi_{\pi}^T(x_3))\nonumber\\
&&+r_{f_0}\Phi_{f_0}^T(x_2)
((x_2+x_3-1)\Phi_{\pi}^P(x_3)+(-x_2+x_3-1)\Phi_{\pi}^T(x_3))
\nonumber\\
 &&+r_{f_0}\Phi_{f_0}^S(x_2)((x_2-x_3+1)\Phi_{\pi}^P(x_3)-(x_2+x_3-1)\Phi_{\pi}^T(x_3))]
\nonumber\\
&&\times E'_{ei}(t)h_n(x_1,1-x_3,x_2,b_1,b_3)+E'_{ei}(t') h_n(x_1,x_3,x_2,b_1,b_3)\nonumber\\
&& \times [-x_3\Phi_{f_0}(x_2)(\Phi^T_{\pi}(x_3)-\Phi_{\pi}^P(x_3))-r_{f_0}x_3(\Phi_{f_0}^S(x_2)-\Phi_{f_0}^T(x_2))
\nonumber\\
&&\times(\Phi_{\pi}^P(x_3)-\Phi_{\pi}^T(x_3))-r_{f_0}x_2
(\Phi_{f_0}^S(x_2)+\Phi_{f_0}^T(x_2))(\Phi_{\pi}^P(x_3)+\Phi_{\pi}^T(x_3))]
 \bigg\}\;,
\end{eqnarray}
\begin{eqnarray}
{\cal M}^{P2}_{ef_0} &=& -32 \pi C_Fm_{B_s}^4 /\sqrt{2N_C}\int_0^1
dx_1dx_2dx_3 \int_0^{\infty} b_1 db_1\, b_3
db_3\,\Phi_{B_s}(x_1,b_1)\Phi_{\pi}^A(x_3)
\nonumber \\
& &\times \bigg\{ \left[ (x_3-x_2-1)\Phi_{f_0}(x_2) -r_{f_0}x_2\left(
\Phi_{f_0}^S(x_2)+\Phi_{f_0}^T(x_2) \right)\right]
\nonumber \\
& & \times E'_{ei}(t)
h_n(x_1,1-x_3,x_2,b_1,b_3)+ E'_{ei}(t')
h_n(x_1,x_3,x_2,b_1,b_3)\nonumber \\
& &\times \left[x_2\Phi_{f_0}(x_2)+r_{f_0}
x_2(\Phi_{f_0}^S(x_2)-\Phi_{f_0}^T(x_2))\right] \bigg\}\;.
\end{eqnarray}

For the non-factorizable annihilation diagrams (e) and (f), again
all three wave functions are involved. For the $(V-A)(V-A)$ and $(S-P)(S+P)$
operators, the results are
\begin{eqnarray}
 {\cal M}_{af_0} &=& -32\pi C_Fm_{B_s}^4 /\sqrt{2N_C}\int_0^1
dx_1dx_2dx_3 \int_0^{\infty} b_1 db_1\, b_3
db_3\,
\nonumber \\
& &\times \Phi_{B_s}(x_1,b_1)\bigg\{ E'_{ai}(t) h_{na}(x_1,x_3,x_2,b_1,b_3)\left[ x_3\Phi_{\pi}^A(x_3)
\Phi_{f_0}(x_2)\right.
\nonumber\\
& &\left.+r_{\pi}r_{f_0} \Phi_{f_0}^T(x_2) (
(x_2-x_3+1)\Phi_{\pi}^T(x_3)-(x_2+x_3-1) \Phi_{\pi}^P(x_3))
\right.\nonumber\\
& &\left.
+ r_{\pi}r_{f_0} \Phi_{f_0}^S(x_2)(
(-x_2+x_3+3)\Phi_{\pi}^P(x_3)+(x_2+x_3-1) \Phi_{\pi}^T(x_3))
\right]
\nonumber \\
& &+  E'_{ai}(t') h'_{na}(x_1,x_3,x_2,b_1,b_3)
\left[
(x_2-1)\Phi_{\pi}^A(x_3) \Phi_{f_0}(x_2)
\right.\nonumber \\
& &\left.+r_{\pi}r_{f_0} \Phi_{f_0}^T(x_2) \left(
(-x_2+x_3+1)\Phi_{\pi}^T(x_3)-(x_2+x_3-1) \Phi_{\pi}^P(x_3)
\right)\right.\nonumber\\
& &\left. + r_{\pi}r_{f_0} \Phi_{f_0}^S(x_2) \left(
(x_2-x_3-1)\Phi_{\pi}^P(x_3)+(x_2+x_3-1) \Phi_{\pi}^T(x_3)
\right)\right] \bigg\}\;,
\end{eqnarray}
\begin{eqnarray}
 {\cal M}^{P2}_{af_0} &=& -32 \pi C_Fm_B^4 /\sqrt{2N_C}\int_0^1
dx_1dx_2dx_3 \int_0^{\infty} b_1 db_1\, b_3 db_3\,\Phi_{B_s}(x_1,b_1)
\nonumber \\
&&\times \bigg\{ \left[(x_2-1)\Phi_{f_0}(x_2)
\Phi^A_{\pi}(x_3)-4r_{\pi}r_{f_0}\Phi^S_{f_0}(x_2)\Phi_{\pi}^P(x_3)
+r_{\pi}r_{f_0}\left((x_3-x_2-1)\right.\right.\nonumber
\\&&\;\left.\left.\times\left(\Phi_{\pi}^P(x_3)\Phi_{f_0}^S(x_2)+\Phi_{\pi}^T(x_3)\Phi_{f_0}^T(x_2)\right)
-(x_2+x_3-1)\left(\Phi_{\pi}^P(x_3)\Phi_{f_0}^T(x_2)\right.\right.\right.\nonumber \\
&&\left.\left.\left.-\Phi_{\pi}^T(x_3)\Phi_{f_0}^S(x_2)\right)\right)\right] E'_{ai}(t) h_{na}(x_1,x_3,x_2,b_1,b_3)
+E'_{ai}(t') h'_{na}(x_1,x_3,x_2,b_1,b_3)
\nonumber \\
&& \;\;\times \left[x_3 \Phi_{f_0}(x_2)
\Phi^A_{\pi}(x_3)+x_3r_{\pi}r_{f_0}(\Phi_{f_0}^S(x2)-\Phi_{f_0}^T(x2))(\Phi_{\pi}^P(x_3)+\Phi_{\pi}^T(x_3))
\right.\nonumber \\ &&\;\;\;\left. +r_{\pi}r_{f_0}
(1-x_2)(\Phi_{f_0}^S(x2)+\Phi_{f_0}^T(x2))(\Phi_{\pi}^P(x_3)-\Phi_{\pi}^T(x_3))
\right]\bigg\}\;.
\end{eqnarray}

The factorizable annihilation diagrams (g) and (h) involve only the
$\pi$ and $f_0$ mesons' wave functions. There are three
kinds of decay amplitudes for these two diagrams. $F_{af_0}$ is for
$(V-A)(V-A)$ type operators, $F_{af_0}^{P1}$ is for $(V-A)(V+A)$
type operators, while $F_{af_0}^{P2}$ is for $(S-P)(S+P)$ type
operators:
\begin{eqnarray}
F_{af_0} &=&F^{P1}_{af_0}= 8 \pi C_F m_{B_s}^4f_{B_s} \int_0^1 dx_2
dx_3 \int_0^{\infty} b_2db_2\, b_3db_3\
\bigg\{\left[(x_2-1)\Phi_{\pi}^A(x_3)\Phi_{f_0}(x_2)
\right.\nonumber \\
&& \left.  + 2r_{\pi}
r_{f_0}(x_2-2) \Phi_{\pi}^P(x_3)\Phi_{f_0}^S(x_2) -2r_{\pi} r_{f_0}x_2
\Phi_{\pi}^P(x_3)\Phi_{f_0}^T(x_2) \right]
\nonumber \\
&&\times E_{ai}(t) h_{a}(x_3,1-x_2, b_3, b_2)+E_{ai}(t') h_{a}(1-x_2,x_3, b_2, b_3)\left[x_3\Phi_{\pi}^A(x_3)\Phi_{f_0}(x_2)
\right.\nonumber\\
&&\left.
+2r_{\pi}
r_{f_0}\Phi_{f_0}^S(x_2)((x_3+1)\Phi_{\pi}^P(x_3)+(x_3-1)\Phi_{\pi}^T(x_3))\right]\bigg\},
\end{eqnarray}
\begin{eqnarray}
F^{P2}_{af_0} &=& 16 \pi C_F m_{B_s}^4f_{B_s} \int_0^1
dx_2 dx_3 \int_0^{\infty} b_2db_2\, b_3db_3\,  \nonumber\\
&&
\times\bigg\{[r_{f_0}(x_2-1)\Phi_{\pi}^A(x_3)(\Phi_{f_0}^S(x_2)+\Phi_{f_0}^T(x_2))
-2r_{\pi}\Phi_{\pi}^P(x_3)\Phi_{f_0}(x_2)]\nonumber\\
 &&\times E_{ai}(t) h_{a}(x_3,1-x_2,
b_2, b_3) - E_{ai}(t') h_{a}(1-x_2,x_3, b_2,
b_3)\nonumber\\
 &&
\times [2r_{f_0}\Phi^A_{\pi}(x_3)
\Phi_{f_0}^S(x_2)+r_{\pi}x_3\Phi_{f_0}(x_2)
(\Phi_{\pi}^P(x_3)-\Phi_{\pi}^T(x_3))]\bigg\}.\end{eqnarray}

If we exchange the $\pi^0$ and $f_0$ in Fig.\ref{Figure1}, the result will be different. In the considered decays, the meson
$f_0$ cannot lie in the emitted position, like $\pi$' position  in Fig.1(a-d). So only the annihilation type diagrams left, just like
Fig.1(e-h), can
give contributions. They are listed as follows:
\be {\cal M}_{a\pi} &=& 32\pi C_Fm_{B_s}^4 /\sqrt{2N_C}\int_0^1
dx_1dx_2dx_3 \int_0^{\infty} b_1 db_1\, b_2
db_2\,\Phi_{B_s}(x_1,b_1)
 \bigg\{ \left[ -x_2\Phi_{\pi}^A(x_3)
\right.\nonumber \\
& &\times  \Phi_{f_0}(x_2)+r_{\pi}r_{f_0} \Phi_{f_0}^T(x_2) \left(
(x_2+x_3-1)\Phi_{\pi}^P(x_3)+(-x_2+x_3+1) \Phi_{\pi}^T(x_3)
\right)\nonumber\\
& &\left. + r_{\pi}r_{f_0} \Phi_{f_0}^S(x_2) \left(
(x_2-x_3+3)\Phi_{\pi}^P(x_3)-(x_2+x_3-1) \Phi_{\pi}^T(x_3)
\right)\nonumber\right]
\nonumber \\
& &\times E'_{ai}(t) h_{na}(x_1,x_2,x_3,b_1,b_2)-  E'_{ai}(t') h'_{na}(x_1,x_2,x_3,b_1,b_2)\left[
(x_3-1)\Phi_{\pi}^A(x_3)
\right.\nonumber \\
& &\times\Phi_{f_0}(x_2)+r_{\pi}r_{f_0} \Phi_{f_0}^S(x_2) \left(
(x_2-x_3+1)\Phi_{\pi}^P(x_3)-(x_2+x_3-1) \Phi_{\pi}^T(x_3)
\right)\nonumber\\
& &\left. + r_{\pi}r_{f_0} \Phi_{f_0}^T(x_2) \left(
(x_2+x_3-1)\Phi_{\pi}^P(x_3)-(1+x_2-x_3) \Phi_{\pi}^T(x_2)
\right)\right] \bigg\}\;, \en
\begin{eqnarray}
 {\cal M}^{P2}_{a\pi} &=& -32 \pi C_Fm_{B_s}^4 /\sqrt{2N_C}\int_0^1
dx_1dx_2dx_3 \int_0^{\infty} b_1 db_1\, b_2
db_2\,\Phi_{B_s}(x_1,b_1)\bigg\{ \left[4r_{\pi}r_{f_0}\Phi^S_{f_0}(x_2)
\right.\nonumber \\
&&\left.\times \Phi_{\pi}^P(x_3)+(x_3-1)\Phi_{f_0}(x_2)
\right.\Phi_{\pi}^A(x_3)
+r_{\pi}r_{f_0}\left((x_2-x_3-1)\left(\Phi_{\pi}^P(x_3)\Phi_{f_0}^S(x_2)\right.\right.\nonumber
\\&&\left.-\Phi_{\pi}^T(x_3)\Phi_{f_0}^T(x_2)\right.)\left.
-(x_2+x_3-1)(\Phi_{\pi}^P(x_3)\Phi_{f_0}^T(x_2)-\Phi_{\pi}^T(x_3)\Phi_{f_0}^S(x_2))\right]\nonumber \\
&&\times E'_{ai}(t) h_{na}(x_1,x_2,x_3,b_1,b_2)+ \left[x_2 \Phi_{f_0}(x_2)
\Phi_{\pi}^A(x_3)-x_2r_{\pi}r_{f_0}(\Phi_{f_0}^S(x2)+\Phi_{f_0}^T(x2))
\right.\nonumber \\
&&\left. \times(\Phi_{\pi}^P(x_3)-\Phi_{\pi}^T(x_3))-r_{\pi}r_{f_0}
(1-x_3)(\Phi_{f_0}^S(x2)-\Phi_{f_0}^T(x2))(\Phi_{\pi}^P(x_3)+\Phi_{\pi}^T(x_3))
\right]\nonumber \\
&&\times E'_{ai}(t') h'_{na}(x_1,x_2,x_3,b_1,b_2) \bigg\}\;,
\end{eqnarray}
\begin{eqnarray}
F_{a\pi} &=&-F^{P1}_{a\pi}= 8 \pi C_F m_{B_s}^4f_{B_s} \int_0^1 dx_2 dx_3
\int_0^{\infty} b_2db_2\, b_3db_3\  \bigg\{\left[(x_3-1)\Phi_{\pi}^A(x_3)\Phi_{f_0}(x_2)
\right.\nonumber \\
&& \left. - 2r_{\pi}
r_{f_0}(x_3-2) \Phi_{\pi}^P(x_3)\Phi_{f_0}^S(x_2) +2r_{\pi} r_{f_0}x_3
\Phi_{\pi}^T(x_3)\Phi_{f_0}^S(x_2) \right]
\nonumber \\
&&\times E_{ai}(t) h_{a}(x_2,1-x_3, b_2, b_3)+E_{ai}(t') h_{a}(1-x_3,x_2, b_3, b_2)
\nonumber\\
&&
 \times [x_2\Phi_{\pi}^A(x_3)\Phi_{f_0}(x_2)-2r_{\pi}
r_{f_0}\Phi_{\pi}^P(x_3)((x_2+1)\Phi_{f_0}^S(x_2)+(x_2-1)\Phi_{f_0}^T(x_2))] \bigg\}, \en

\be
 F^{P2}_{a\pi} &=& -16 \pi C_F m_{B_s}^4f_{B_s} \int_0^1 dx_2
dx_3 \int_0^{\infty} b_2db_2\, b_3db_3\,   \nonumber \\&&
\times\bigg\{[r_{\pi}(x_3-1)\Phi_{f_0}(x_2)(\Phi_{\pi}^P(x_3)+\Phi_{\pi}^T(x_3))
+2r_{f_0}\Phi_{\pi}(x_3)\Phi_{f_0}^S(x_2)] \nonumber \\&&\times E_{ai}(t) h_{a}(x_2,1-x_3,
b_2, b_3) -E_{ai}(t') h_{a}(1-x_3,x_2,
b_3, b_2)\nonumber \\&&\times [2r_{\pi}\Phi_{\pi}^P(x_3)
\Phi_{f_0}(x_2)+r_{f_0}x_2\Phi^A_{\pi}(x_3)
(\Phi_{f_0}^T(x_2)-\Phi_{f_0}^S(x_2))]\bigg\}\;.\en

In the above formulae, the function $E$ are defined as:
\begin{eqnarray}
E_{ei}(t)&=&\alpha_s(t)\,\exp[-S_B(t)-S_{3}(t)],\\
E'_{ei}(t)&=&\alpha_s(t)\,\exp[-S_B(t)-S_{2}(t)-S_{3}(t)]|_{b_3=b_1},\\
E_{ai}(t)&=&\alpha_s(t)\,\exp[-S_2(t)-S_{3}(t)],\\
E'_{ai}(t)&=&\alpha_s(t)\,\exp[-S_B(t)-S_{2}(t)-S_{3}(t)]|_{b_3=b_2},
\end{eqnarray}
where $\alpha_s$ is the strong coupling constant, $S$ is the Sudakov form factor.
In our numerical analysis, we use the one-loop expression for the
strong coupling constant; we use  $c=0.4$ for the parameter in the
jet function. The explicit form of $h$ and $S$ have been given in
\cite{PQCD}.

Combining the contributions from different diagrams, the total decay
amplitudes for these decays can be written as: \be
{\cal M} (\bar B^0_s\to f_0\pi) &=&{\cal
M}_{s\bar{s}}\times\cos\theta+\frac{1}{\sqrt{2}}{\cal M}_{n\bar
n}\sin\theta, \en
where $\theta$ is mixing angle, and
\be {\cal M}_{s\bar{s}}&=&V_{ub}V^*_{us}(F_{ef_0}a_2+M_{ef_0}C_2)
-\frac{3}{2}V_{tb}V^*_{ts}\left[F^{P2}_{ef_0}(a_9-a_{7})
+M^{P1}_{ef_0}C_{10}+M^{P2}_{ef_0}C_{8} \right] , \label{ssbar}\en
\be {\cal M}_{n\bar{n}}&=&
V_{ub}V^*_{us}\left[(M_{a\pi}+M_{af_0})C_2+(F_{a\pi}+F_{af_0})a_2\right]-
\frac{3}{2}V_{tb}V^*_{ts}\left[(M_{a\pi}+M_{af_0})C_{10}
\right.\nonumber\\
&&\left.+(M^{P2}_{a\pi}+M^{P2}_{af_0})C_{8}+
(F_{a\pi}+F_{af_0})(a_9-a_{7})\right]
, \label{nnbar}\en
where the combinations of the Wilson coefficients are defined as usual
\cite{AKL,keta}:
\begin{eqnarray}
a_1= C_2+C_1/3, & a_3= C_3+C_4/3,~a_5= C_5+C_6/3,~a_7=
C_7+C_8/3,~a_9= C_9+C_{10}/3,\quad\quad\\
a_2= C_1+C_2/3, & a_4= C_4+C_3/3,~a_6= C_6+C_5/3,~a_8=
C_8+C_7/3,~a_{10}= C_{10}+C_{9}/3.\quad\quad
\end{eqnarray}

\section{Numerical results and discussions}
The twist-2 LCDA $\Phi_{f_0}$ can be expanded in the Gegenbauer polynomials: \be
\Phi_{f_0}(x,\mu)&=&\frac{1}{2\sqrt{2N_c}}\bar
f_{f_0}(\mu)6x(1-x)\sum_{m=1}^\infty B_m(\mu)C^{3/2}_m(2x-1),
\en
where $B_m(\mu)$ and $C^{3/2}_m(x)$ are the Gegenbauer moments and Gegenbauer polynomials, respectively. The values for Gegenbauer moments
and the decay constants are taken (at scale $\mu=1 \mbox{GeV}$) as:
\be
\mbox {Scenario I}: \bar f_{f_0(980)}&=&(0.37\pm0.02) \mbox {GeV}, \quad\bar f_{f_0(1500)}=-(0.255\pm0.03) \mbox {GeV},\non
B_1(980)&=&-0.78\pm0.08, \quad\quad\quad B_3(980)=0.02\pm0.07,\non
B_1(1500)&=&0.80\pm0.40, \quad\quad\;\quad B_3(1500)=-1.32\pm0.14;\non
\mbox{Scenario II}: \bar f_{f_0(1500)}&=&(0.49\pm0.05) \mbox {GeV},\non
B_1(1500)&=&-0.48\pm0.11,\quad\quad\quad B_3(1500)=-0.37\pm0.20.
\en

As for the explicit form of the Gegenbauer moments for the twist-3
distribution amplitudes $\Phi_{f_0}^S$ and $\Phi_{f_0}^T$, although they have
been studied in the Ref. \cite{cdlv}, we adopt the asymptotic form: \be
\Phi^S_{f_0}&=& \frac{1}{2\sqrt {2N_c}}\bar f_{f_0},\,\,\,\,\,\,\,\Phi_{f_0}^T=
\frac{1}{2\sqrt {2N_c}}\bar f_{f_0}(1-2x). \en

The twist-2 pion distribution
amplitude $\Phi^{A}_{\pi}$, and the twist-3 ones $\Phi^{P}_{\pi}$ and
$\Phi^{T}_{\pi}$ have been parametrized as \cite{pball}:
\begin{eqnarray}
 \Phi_{\pi}^{A}(x) &=&  \frac{f_{\pi}}{2\sqrt{2N_c} }
    6x (1-x)
    \left[1+0.17(5(1-2x)^2-1)\right.\non &&\left.\qquad\qquad-0.028(1-14(1-2x)^2+21(1-2x)^4)
  \right],\label{piw1}\\
 \Phi^{P}_{\pi}(x) &=&   \frac{f_\pi}{2\sqrt{2N_c} }
   \left[ 1+0.21(3(1-2x)^2-1)\right.\non &&\left.\qquad\qquad-0.11/8(3-30(1-2x)^2+35(1-2x)^4)\right]  ,\\
 \Phi^{T}_{\pi}(x) &=&  \frac{f_\pi}{2\sqrt{2N_c} } (1-2x)
   \left[ 1+0.56
   (1-10x+10x^2)\right] .\quad\quad\label{piw}
 \end{eqnarray}
The $B_s$ meson's wave function can be written as:
\be
\phi_{B_s}(x,b)=N_{B_s}x^2(1-x)^2\exp[-\frac{M^2_{B_s}x^2}{2\omega^2_{b_s}}-\frac{1}{2}(\omega_{b_s}b)^2],
\en
where $\omega_{b_s}$ is a free parameter and we take $\omega_{b_s}=0.5\pm0.05$ GeV in numerical calculations,
and $N_{B_s}=63.67$ is the normalization factor for $\omega_{b_s}=0.5$.

For the numerical calculation, we list the other input parameters in Table
~I.
\begin{table}\caption{Input parameters used in the numerical calculation\cite{CCYscalar,pdg08}. }
\begin{center}
\begin{tabular}{c |cc}
\hline \hline
 Masses &$m_{f_0(980)}=0.98 \mbox{ GeV}$,   &$ m_0^{\pi}=1.3 \mbox{ GeV}$, \\ \
  & $m_{f_0(1500)}=1.5 \mbox{ GeV}$&$ M_{B_s} = 5.37 \mbox{ GeV}$,\\
 \hline
  Decay constants &$f_{B_s} = 0.23 \mbox{ GeV}$,  & $ f_{\pi} = 0.13
 \mbox{ GeV}$,\\
 \hline
Lifetimes &
$\tau_{B^0_s}=1.466\times 10^{-12}\mbox{ s}$,&\\
 \hline
$CKM$ &$V_{tb}=1.0$, & $V_{ts}=0.387$,\\
 &$V_{us}=0.2255$, & $V_{ub}=0.00393e^{-i60^{\circ}}$.\\
\hline \hline
\end{tabular}\label{para}
\end{center}
\end{table}

If $f_0(980)$ and $f_0(1500)$ are purely composed of $n\bar n$($s\bar s$), the branching ratios
of $\bar B^0_s\to f_0(980)\pi^0, f_0(1500)\pi^0$ are:
\be
{\cal B}(\bar B^0_s\to
f_0(980)(n\bar n)\pi^0)&=&(0.46^{+0.7+1.2+0.1}_{-0.5-1.0-0.0})\times 10^{-8}, \mbox{ Scenario I},\\
{\cal B}(\bar B^0_s\to
f_0(1500)(n\bar n)\pi^0)&=&(2.4^{+0.1+3.4+0.5}_{-0.1-2.0-0.3})\times 10^{-8}, \mbox{ Scenario I},\\
{\cal B}(\bar B^0_s\to
f_0(1500)(n\bar n)\pi^0)&=&(1.1^{+0.8+1.4+0.2}_{-0.7-1.3-0.1})\times 10^{-8}, \mbox{ Scenario II};\\
{\cal B}(\bar B^0_s\to
f_0(980)(s\bar s)\pi^0)&=&(1.7^{+0.1+0.1+0.1}_{-0.1-0.2-0.1})\times 10^{-7}, \mbox{ Scenario I},\\
{\cal B}(\bar B^0_s\to
f_0(1500)(s\bar s)\pi^0)&=&(0.66^{+0.00+0.30+0.06}_{-0.00-0.13-0.05})\times 10^{-7}, \mbox{ Scenario I},\\
{\cal B}(\bar B^0_s\to
f_0(1500)(s\bar s)\pi^0)&=&(3.8^{+0.5+0.5+0.7}_{-0.4-0.4-0.5})\times 10^{-7}, \mbox{ Scenario II},
\end{eqnarray}
where the uncertainties are from the decay constant of $f_0$,
the Gegenbauer moments $B_1$ and $B_3$. One can see that the values of
${\cal B}(\bar B_s\to f_0(n\bar n)\pi^0)$ are smaller than the corresponding values of ${\cal B}(\bar B_s\to f_0(s\bar s)\pi^0)$,
it is contrary
to the case of $\bar B_s\to f_0(980)K^0, f_0(1500)K^0$ decays \cite{zqzhang3}.

\begin{figure}[t,b]
\begin{center}
\includegraphics[scale=0.7]{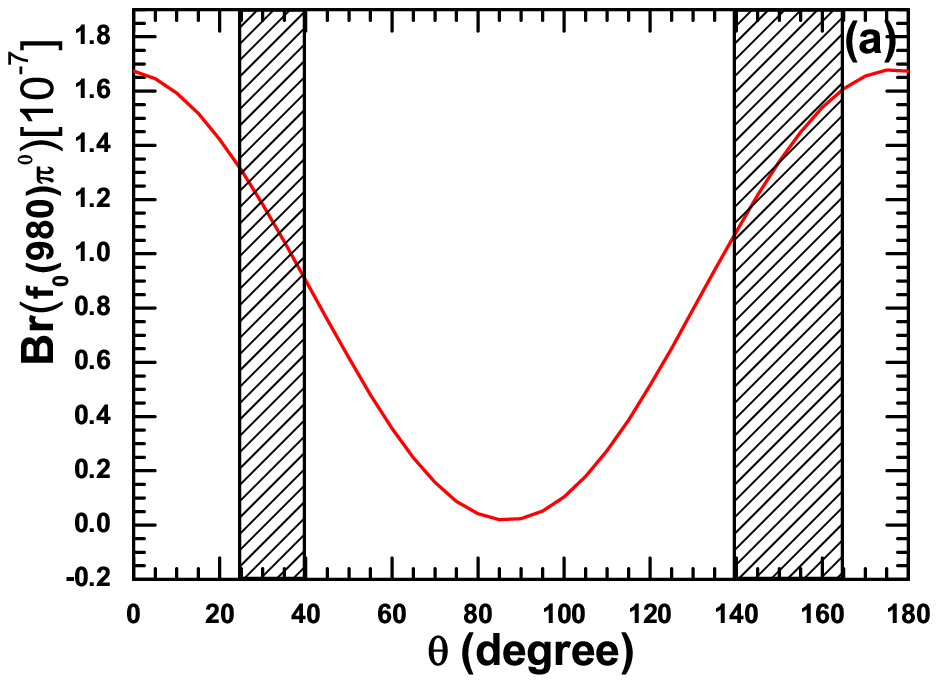}
\includegraphics[scale=0.7]{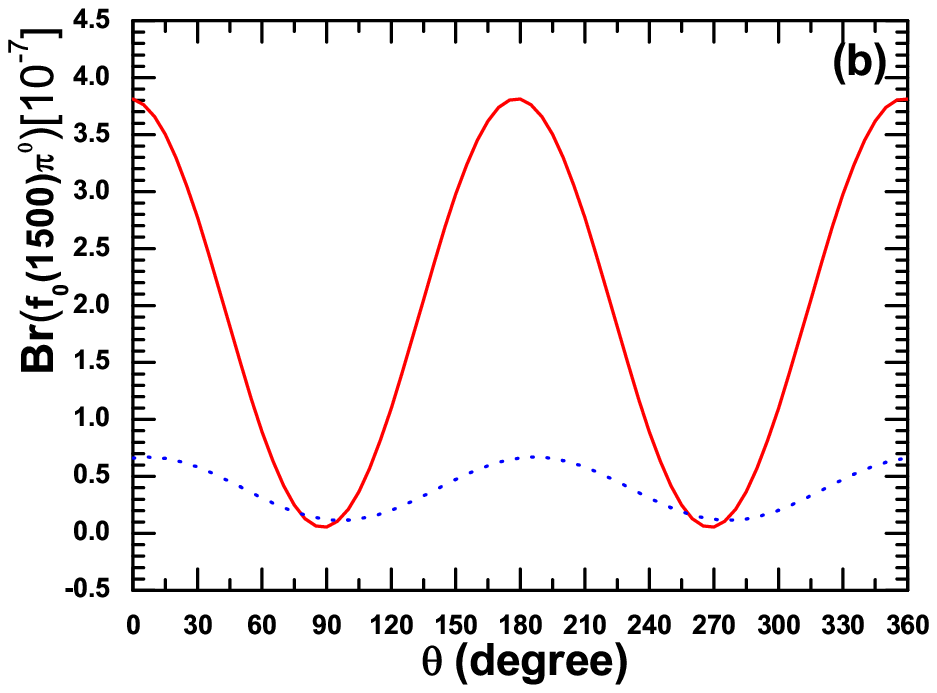}
\vspace{0.3cm} \caption{The dependence of the branching ratios for
$\bar B^0_s\to f_0(980)\pi^0$ (a) and $\bar B^0_s\to f_0(1500)\pi^0$ (b) on the
mixing angle $\theta$ using the inputs derived from QCD sum rules.
The vertical bands show
two possible ranges of $\theta$: $25^\circ<\theta<40^\circ$ and $140^\circ<\theta<165^\circ$.
For the right panel, the solid (dotted) curve is plotted in scenario II (I).}\label{fig2}
\end{center}
\end{figure}
\begin{table}
\caption{ Decay amplitudes for $\bar B^0_s\to f_0(980)\pi^0,f_0(1500)\pi^0$ ($\times 10^{-2} \mbox {GeV}^3$).}
\begin{center}
\begin{tabular}{cc|c|c|c|c|c}
\hline \hline  $\bar ss$&&$F^{\pi,T}_e$ &$F^{\pi}_e$ & $M^{\pi,T}_e$ & $M^{\pi}_e$ \\
\hline
$f_0(980)\pi^0$ (SI) &   &8.98&1.50&$-3.51+5.75i$&$-0.01+0.04i$ \\
$f_0(1500)\pi^0$ (SI) &   &-11.9&-1.03&$-5.89+4.70i$&$-0.02+0.04i$ \\
$f_0(1500)\pi^0$ (SII) &   &17.8&2.38&$-0.77+3.64i$& $0.001+0.02i$\\
\hline \hline
  $\bar nn $&&$M^{f_0,T}_a+M^{\pi,T}_a$ &  $M^{f_0}_a+M^{\pi}_a$ & $F^{f_0,T}_a+F^{\pi,T}_a$ & $F^{f_0}_a+F^\pi_a$  \\
\hline
$f_0(980)(\bar nn)\pi^0$ (SI)& &$6.3+6.9i$&$-0.05+0.03i$& $0.89-0.11i$&$-0.001+0.025i$\\
$f_0(1500)(\bar nn)\pi^0$ (SI)& &$-13.8+20.8i$&$0.11+0.03i$&$-0.015+1.34i$ &$-0.02-0.017i$\\
$f_0(1500)(\bar nn)\pi^0$ (SII)& &$15.3-3.1i$&$0.01-0.01i$&$1.62-0.63i$ &$0.04+0.04i$\\
\hline \hline
\end{tabular}\label{amp}
\end{center}
\end{table}

In Table II, we list values of the factorizable and non-factorizable amplitudes from the emission and annihilation topology
diagrams of $\bar B^0_s\to f_0(980)\pi^0, f_0(1500)\pi^0$. $F^{\pi}_{e(a)}$ and $M^{\pi}_{e(a)}$ are the $\pi^0$ emission (annihilation)
factorizable
contributions and non-factorizable contributions from penguin operators respectively. Similarly, $F^{f_0}_{e(a)}$ and $M^{f_0}_{e(a)}$ denote as
the contributions from $f_0$ emission (annihilation) factorizable contributions and non-factorizable contributions from penguin operators
respectively. $F^{\pi(f_0),T}_{e(a)}, M^{\pi(f_0),T}_{e(a)}$ denote the corresponding contributions from tree operators
$O_2, O_1$. From the table, regardless of the CKM suppression, one can find that
the contributions from tree operators are (much) larger than the corresponding ones from penguin operators, especially
for the non-factorizable emission diagrams and the annihilation diagrams. In fact, the tree operators contributions
are strongly CKM-suppressed compared to penguin operators contributions. Here considering the amplitudes of
annihilation diagrams is necessary.

In Fig.~\ref{fig2}, we plot the branching ratios as functions of the
mixing angle $\theta$. Because there are many discrepancies about
the mixing of quark components for the meson $f_0(1500)$, we show
the dependence of the branching ratio for $\bar B^0_s\to
f_0(1500)\pi^0$ on all the mixing angle values, that is $(0^\circ,
360^\circ)$. In the allowed mixing angle ranges, the branching ratio
of $\bar B^0_s\to f_0(980)\pi^0$ is: \be 1.0\times 10^{-7}<{\cal
B}(\bar B^0_s\to f_0(980)\pi^0)<1.6\times 10^{-7}, \en which is
smaller than the branching ratio of $\bar B^0_s\to f_0(980)K^0$. The
difference is a few times even one order. As to the decay $\bar
B^0_s\to f_0(1500)\pi^0$, it is interesting that this channel is
better to distinguish between the first excited state (scenario I) and the
lowest lying state (scenario II) for $f_0(1500)$. The branching ratio of
$\bar B^0_s\to f_0(1500)\pi^0$ for scenario I is at the order of
$10^{-8}$ in $(0^\circ,360^\circ)$, while its value for scenario II
is at the order of $10^{-7}$ in most of mixing angle ranges, except
for $(60^\circ,120^\circ)$ and $(240^\circ,300^\circ)$. If the
mixing angle is not in these two ranges, it is ease to determine the
nature of $f_0(1500)$. If the observation of the decay $\bar
B^0_s\to f_0(1500)\pi^0$ at the order of $10^{-7}$, it would
indicate that scenario II is favored. We also find the branching
ratios of these decays are smaller than those of corresponding
channels $\bar B^0_s\to f_0(980)K^0, f_0(1500)K^0$ about a few times
even one order.

Now we turn to the evaluations of the CP-violating asymmetries of
the considered decays in the PQCD approach.
For the neutral decays $\bar B^0_s\to f_0(980)\pi^0,f_0(1500)\pi^0$,
there are both direct $CP$ asymmetry $A^{dir}_{CP}$ and
mixing-induced $CP$ asymmetry $A^{mix}_{CP}$. The time dependent
$CP$ asymmetry of $B_s$ decay into a $CP$ eigenstate $f$ is defined
as:
\be
{\cal A}_{CP}(t)={\cal A}^{dir}_{CP}(B_s\to f)\cos(\Delta m_s)+{\cal A}^{mix}_{CP}(B_s\to f)\sin(\Delta m_s),
\en
with
\be
{\cal A} ^{dir}_{CP}(B_s\to f)&=&\frac{|\lambda|^2-1}{1+|\lambda|^2},
\;\;\;
{\cal A}^{mix}_{CP}(B_s\to f)=\frac{2 Im \lambda}{1+|\lambda|^2}, \label{dircp}\\
\lambda&=&\eta e^{-2i\beta}\frac{{\cal A}(\bar B_s\to f)}{{\cal
A}(B_s\to f),} \label{lambda}
\en
where  $\eta=\pm1$ depends on the $CP$ eigenvalue
of $f$, $\Delta m_s$ is the mass difference of the two neutral $B_s$
meson eigenstates. Here we only give the direct CP-violating asymmetry.

\begin{table}
\caption{Direct $CP$ asymmetries (in units of \%) of $\bar B^0_s \to f_0(980)\pi^0, f_0(1500)\pi^0$ decays for $n\bar n$
and $s\bar s$ components, respectively.}\label{CP}
\begin{center}
\begin{tabular}{c|c|c|c}
   \hline \hline
   Channel & Scenario I & Scenario II &  \\
   \hline   $\bar B^0_s \to f_0(980)(n\bar n) \pi^0$&$51.5$&-\\
    $\bar B^0_s \to f_0(980)(s\bar s) \pi^0$&$-16.0$&-\\
 \hline     $\bar B^0_s \to f_0(1500)(n\bar n) \pi^0 $&$29.0$&$15.1$\\
    $\bar B^0_s \to f_0(1500)(s\bar s)\pi^0 $&$19.3$&$-6.7$\\
   \hline\hline
\end{tabular}
   \end{center}
\end{table}

\begin{figure}[t,b]
\begin{center}
\includegraphics[scale=0.7]{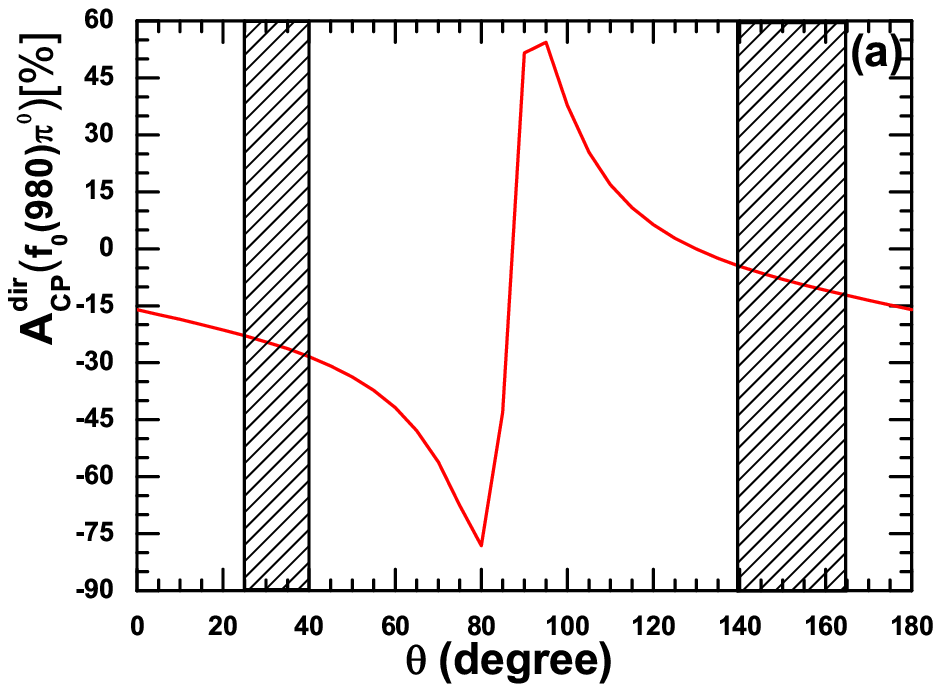}
\includegraphics[scale=0.7]{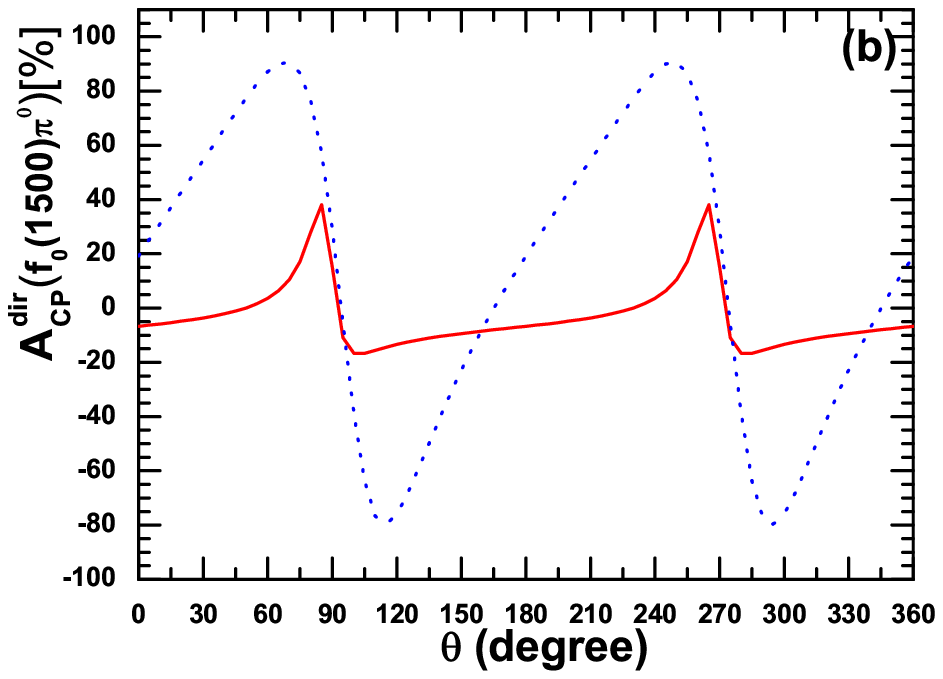}
\vspace{0.3cm} \caption{The dependence of the direct CP asymmetries for
$\bar B^0_s\to f_0(980)\pi^0$ (a) and $\bar B^0_s\to f_0(1500)\pi^0$ (b) on the
mixing angle $\theta$. For the right panel, the solid (dotted) curve is plotted in scenario II (I).
The vertical bands show two possible ranges of $\theta$: $25^\circ<\theta<40^\circ$ and $140^\circ<\theta<165^\circ$.}\label{fig3}
\end{center}
\end{figure}

The direct CP asymmetries of $\bar B^0_s\to f_0(n\bar n)\pi^0, f_0(s\bar s)\pi^0$
are listed in Table III. From the definition of the direct CP asymmetry Eq.(\ref{dircp}) and Eq.(\ref{lambda}), we can find the sign of
${\cal A} ^{dir}_{CP}$ is determined by the formula:
\begin{eqnarray}
M^T_rM^P_i-M^T_iM^P_r=\begin{cases}
\left(F^{\pi,T}_e+M^{\pi,T}_{e,r}\right)M^\pi_{e,i}-M^{\pi,T}_{e,i}\left(F^{\pi}_e+M^\pi_{e,r}\right), \qquad\text{for} f_0(s\bar s);\\
\left(M^{f_0,T}_{a,r}+M^{\pi,T}_{a,r}+F^{f_0,T}_{a,r}+F^{\pi,T}_{a,r}\right)\left(M^{f_0}_{a,i}+M^{\pi}_{a,i}+F^{f_0}_{a,i}+F^{\pi}_{a,i}\right)
-\left(M^{f_0,T}_{a,i}\right.\nonumber \\\left.+M^{\pi,T}_{a,i}+F^{f_0,T}_{a,i}+F^{\pi,T}_{a,i}\right)\left(M^{f_0}_{a,r}+M^{\pi}_{a,r}+F^{f_0}_{a,r}
+F^{\pi}_{a,r}\right), \quad\text{for} f_0(n\bar n).
\end{cases}
\end{eqnarray}
If $M^T_rM^P_i-M^T_iM^P_r>0$, the sign of the corresponding direct CP asymmetry is positive, contrarily, the value of ${\cal A} ^{dir}_{CP}$
is minus. So one can understand that though the penguin operators contributions are much smaller than the tree operators contributions (the
difference is about two or three orders, seen in Table II), they are important to determine the direct CP asymmetry.
From Table III, it is found that
${\cal A} ^{dir}_{CP}(\bar B^0_s\to f_0(1500)(\bar s s)\pi^0)> 0$ in scenario I, which is contrary to the sign in scenario II.
The direct reason is that there exists an opposite sign for the $\pi^0$ emission
factorizable contributions from penguin operators between these two scenarios.
We also find the direct CP asymmetries for $n\bar n$ components of $f_0(980)$
and $f_0(1500)$ are larger than those  for $s\bar s$ components.

In Fig.3, we plot the direct CP asymmetries of the decays $\bar B^0_s\to f_0(980)\pi^0$ and
$\bar B^0_s\to f_0(1500)\pi^0$. One can see the direct CP asymmetry of $\bar B^0_s\to f_0(980)\pi^0$ is:
\be
-30\%<{\cal A} ^{dir}_{CP}(\bar B^0_s\to f_0(980)\pi^0)<-25\%,
\en
for the mixing angle $\theta$ in the range of $25^\circ<\theta<40^\circ$. While if the mixing angle $\theta$ is taken in the range
$(140^\circ,165^\circ)$, the value of ${\cal A} ^{dir}_{CP}(\bar B^0_s\to f_0(980)\pi^0)$ is about $(-12\sim-5)\%$. For the decay
$\bar B^0_s\to f_0(1500)\pi^0$, if the parameters in scenario II are used, one can find the variation range of
${\cal A} ^{dir}_{CP}(\bar B^0_s\to f_0(1500)\pi^0)$ according to most of the mixing angles is very small, except for the values for mixing angles
near $90^\circ$ or $270^\circ$.
while in
scenario I, the variation range of ${\cal A} ^{dir}_{CP}(\bar B^0_s\to f_0(1500)\pi^0)$ is very large.
The great differences of decay constant and Gegenbauer moments of $f_0(1500)$ result that there exists
great difference for the direct CP asymmetries in two scenarios. It gives the hint that one can determine the
nature of the
meson $f_0(1500)$ by comparing with the future experimental values for these direct CP-violating asymmetries.

It is noticed we consider that the meson $f_0(1500)$ is dominated by
the quarkonium content, the detail discussion can be found in
\cite{zqzhang3}. If we take the mixing mechanism for $f_0(1500)$ as
$\mid f_0(1500)\rangle=-0.54\mid \bar n n\rangle+0.84\mid \bar s
s\rangle+0.03\mid G\rangle$ \cite{mixing} and neglect the small
component of glueball, we have: \be {\cal B}(\bar B^0_s\to
f_0(1500)\pi^0)&=&(4.46^{+0.11+2.47+0.68}_{-0.10-1.85-0.63})\times10^{-8}, \mbox{ Scenario I},\non {\cal
B}(\bar B^0_s\to f_0(1500)\pi^0)&=&(2.81^{+0.61+0.32+0.47}_{-0.54-0.29-0.42})\times10^{-7}, \mbox{
Scenario II};\non {\cal A} ^{dir}_{CP}(\bar B^0_s\to
f_0(1500)\pi^0)&=&-(27.5^{+0.0+0.2+4.1}_{-0.0-0.0-3.3})\%, \qquad\mbox{ Scenario I},\non {\cal A}
^{dir}_{CP}(\bar B^0_s\to f_0(1500)\pi^0)&=&-(9.7^{+0.0+1.2+5.8}_{-0.0-1.1-4.9})\%,  \qquad\mbox{
Scenario II}, \en
which are the values corresponding to the mixing angle $327.3^\circ$. The uncertainties are from the decay constant of $f_0$,
the Gegenbauer moments $B_1$ and $B_3$ for twist-2 LCDAs of the scalar mesons. Certainly, it is only the leading order results.
In this process, the $\pi^0$ emission factorizable amplitudes from the tree operators correspond to the color-suppressed tree
amplitudes, which are known to be modified by the inclusion of the next-to-leading-order (NLO) corrections.
From the calculations of the partial NLO
corrections \cite{keta}, our argument is that the NLO contributions might have a
small influence on the branching ratio. But
it is difficult to say that the predicted discrepancy in the CP
asymmetries must hold under all of the NLO corrections.

\section{Conclusion}\label{summary}

In this paper, we study $ \bar B^0_s\to f_0(980)\pi^0, f_0(1500)\pi^0$ decays
in the PQCD factorization approach and calculate their branching ratios and the direct CP-violating
asymmetries. Several remarks are in order:
\begin{itemize}
\item
If $f_0(980)$ and $f_0(1500)$ are purely composed of $n\bar n$ or $s\bar s$,  one can see that the values
of ${\cal B}(\bar B_s\to f_0(n\bar n)\pi^0)$ are smaller than those of ${\cal B}(\bar B_s\to f_0(s\bar s)\pi^0)$, it is contrary
to the $\bar B_s\to f_0(980)K^0, f_0(1500)K^0$ decays.
\item
In the allowed mixing angle range, the branching ratio of
$\bar B^0_s\to f_0(980)\pi^0$ is:
\be
1.0\times 10^{-7}<{\cal B}(\bar B^0_s\to f_0(980)\pi^0)<1.6\times 10^{-7},
\en
which is smaller than that of the decay $\bar B^0_s\to f_0(980)K^0$. The difference is a few times even one order.
\item The decay $\bar B^0_s\to f_0(1500)\pi^0$ is better to distinguish between
the lowest lying state or the first excited state for $f_0(1500)$. Because its branching ratios for the
two scenarios have about one order difference in most of the mixing angle ranges.  For example, if we take the
mixing mechanism for $f_0(1500)$ as $\mid f_0(1500)\rangle=-0.54\mid \bar n n\rangle+0.84\mid \bar s
s\rangle$, which corresponds to the mixing angle taking about $327.3^\circ$, one can find
\be {\cal B}(\bar B^0_s\to
f_0(1500)\pi^0)&=&4.46\times10^{-8}, \mbox{ Scenario I},\non {\cal
B}(\bar B^0_s\to f_0(1500)\pi^0)&=&2.81\times10^{-7}, \mbox{
Scenario II}.\en
\item
There also exists great difference for the direct CP asymmetries of the decay $\bar B^0_s\to f_0(1500)\pi^0$ in two scenarios.
If the parameters in scenario II are used, one can find the variation range of the value
${\cal A} ^{dir}_{CP}(\bar B^0_s\to f_0(1500)\pi^0)$ according to most of the mixing angles is very small,
except for the values corresponding to mixing angles being
near $90^\circ$ or $270^\circ$, while in
scenario I, the variation range of ${\cal A} ^{dir}_{CP}(\bar B^0_s\to f_0(1500)\pi^0)$ is very large. Certainly,
the NLO contributions may give these direct CP asymmetries some corrections.
\end{itemize}

\section*{Acknowledgment}
This work is partly supported by Foundation of Henan University of Technology under Grant No.150374. The author would like to thank
Hai-Yang Cheng, Cai-Dian L\"U, Wei Wang, Yu-Ming Wang for helpful discussions.

\end{document}